\documentclass[12pt]{elsarticle}  
\usepackage{epsfig,xfrac,subfigure,booktabs,xfrac,amsmath,graphicx}

\newcommand\sr{\mathrm{sr}}
\newcommand\ini{\mathrm{ini}}
\newcommand\en{\mathrm{end}}
\newcommand\diff{\mathrm{d}}
\newcommand\sca{\mathrm{S}}

\newcommand\ret{\mathrm{ret}}
\newcommand\D{\mathrm{d}}
\newcommand\ele{\mathrm{l}}
\newcommand\ere{\mathrm{r}}
\newcommand\e{\mathrm{e}}

\newcommand\phai{\mathrm{pi}}
\newcommand\ua{\mathrm{ua}}
\newcommand\im{\mathrm{i}}

\begin{document}

\title{Semiclassical analysis of the Starobinsky inflationary model}

\author{Truman Tapia$^1$,  Muhammad Zahid Mughal$^2$, Clara Rojas$^3$}

\address{$^1$Yachay Tech University, School of Physical Sciences and Nanotechnology, Hda. San Jos\'e s/n y Proyecto Yachay, 100119, Urcuqu\'i, Ecuador}

\address{$^2$ University of GUJRAT, Department of mathematics,  Jalalpur Jattan Road Gujrat, Pakistan}

\address{$^3$Yachay Tech University, School of Physical Sciences and Nanotechnology, Hda. San Jos\'e s/n y Proyecto Yachay, 100119, Urcuqu\'i, Ecuador}
\ead{crojas@yachaytech.edu.ec}

\begin{abstract}
	In this work we study  the scalar power spectrum and the spectral index for the Starobinsky inflationary model using  the phase integral method up-to  third-order of approximation. We show that the semiclassical methods reproduce the scalar power spectrum for the Starobinsky model with a good accuracy, and the value of the spectral index  compares favorably with observations. Also, we compare the results with the uniform approximation method and the second-order slow-roll approximation.

\end{abstract}

\maketitle

\section{Introduction}

Historically, inflation was introduced to solve the fine-tuning problems \cite{guth:1981}. However, it has another important motivation, this theory predicts the emergence of curvature perturbations with an almost  scale invariant power spectrum, which causes the Cosmic Microwave Background (CMB) anisotropies and the large scale structure of the universe \cite{guth:1982}.  The CMB  anisotropies  allow us to probe the validity of any model of inflation through the comparison of the predicted power spectrum with observations. The anisotropies of the CMB allow us to probe the primordial power spectrum generated in an epoch of cosmological inflation.

Inflation is defined as a period of evolution of the Universe where the expansion was accelerated, so $\ddot{a}(t)>0$, being $a(t)$  the scale factor.
The fundamental characteristic of this theory is a period of expansion extremely fast in a very short period of time that happened when the Universe was extremely young \cite{liddle:2000}. The condition for inflation also can be written as $p<-\frac{\rho}{3}$, then to produce inflation we need matter with the property of having negative pressure. The matter with this property is a scalar field $\phi$, called inflaton. Considering that the dynamics of the inflation field is dictated by a certain potential, which is different for each model of inflation, when the potential  $V(\phi)$ dominates over the kinetic term  $\dot{\phi}^2$ we have inflation. 

The study of plenty models of inflation has been subject to research in the last few decades \cite{martin:2014}. According to the recent results reported by the satellite Planck \cite{akrami:2018}, the Starobinsky potential is  an  inflationary model supported by observations.
This model was introduced in the eighties  \cite{barrow:1988a,barrow:1988b,starobinsky:1980} and has been the cause of interest in recent years \cite{linde:2014,diValentino:2017,paliathanasis:2017,adam:2019,granada:2019,samart:2019,chowdhury:2019}.

The standard method for studying the scalar power spectrum is the slow-roll approximation. 
Another way is to solve numerically the perturbation equation (i.e. Mukhanov-Sasaki equation).  In recent years, semi-classical methods have appeared in the literature as an alternative way to study the equation of perturbations and calculate the power spectrum  \cite{habib:2002,casadio:2005,casadio:2006,rojas:2007b,rojas:2007c,rojas:2009,rojas:2012}. The calculation of the scalar power spectrum  with the uniform approximation  method or the phase-integral method  are faster than the calculations made with a numerical code.

The article is structured as follows: In Sec. II  we show the Starobinsky potential. In Sec. III we show the equations of motion of the Universe and present their solutions numerically and with the slow-roll approximation. In section IV we present the equation of perturbations and its solution with the uniform approximation method and with the phase-integral method up to third-order of approximation to obtain the scalar power spectrum. Also we present the calculation of the scalar power spectrum inside the second-order slow-roll approximation. In section V, we compare the calculation of the scalar power spectrum  obtained in section IV with numerical calculation. In Sec. VI we summarize our results. Finally, in Sec. VII we express our acknowledgment.

\section{Starobinsky potential}

\begin{figure}[th!]
\includegraphics[scale=0.35]{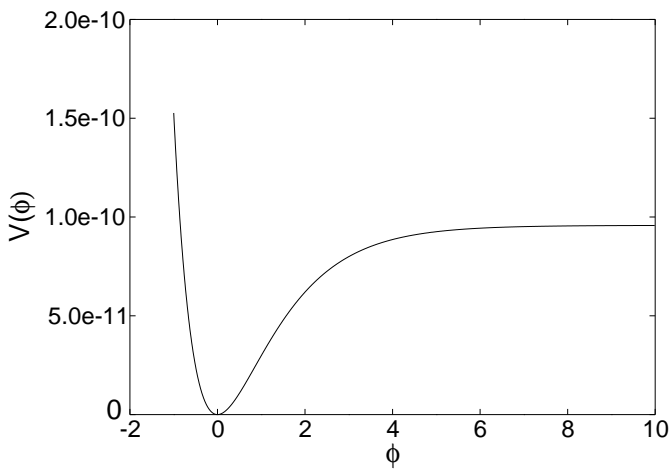}
\caption{Starobinsky potential.}
\label{potential}
\end{figure}

The Starobinsky potential is given by \cite{martin:2014,martin:2019}

\begin{equation}
\label{V}
V(\phi)=M^4 \left(1-e^{-\sqrt{\sfrac{2}{3}}\,\phi}\right)^2,
\end{equation}
where $M=\dfrac{\sqrt{3}}{2} 1.13 \times 10^{-5}$ \cite{mishra:2018}.
In Fig. \ref{potential} we show the Starobinsky potential, at first sight  the potential is sufficiently flat to produce inflation.

\bigskip

\section{Equations of motion}

The equation of motion for a single scalar field are given by the Friedmann equation and the continuity equation

\begin{eqnarray}
\label{Friedmann}
H^2 &=& \dfrac{1}{3} \left[V(\phi) + \dfrac{1}{2} \dot{\phi}^2 \right],\\
\label{continuity}
\ddot{\phi} + 3 H \dot{\phi} &=& -V_{,\phi},
\end{eqnarray}
where the dots indicate derivatives with respect to physical time $t$.

For the Starobinsky inflationary model the Eqs. \eqref{Friedmann} and \eqref{continuity} are not exactly solvable in closed form, they can be solved numerically or using the slow-roll approximation. Using the slow-roll approximation, $\dot{\phi}^2 < V(\phi)$, equations \eqref{Friedmann}  and \eqref{continuity}  becomes \cite{liddle:2000}

\begin{eqnarray}
\label{Friedmannsr}
H^2 &\simeq& \dfrac{1}{3} V(\phi),\\
\label{continuitysr}
3 H \dot{\phi} &\simeq& -V_{,\phi}.
\end{eqnarray}

The slow-roll parameters are given by \cite{liddle:2000},

\begin{eqnarray}
\label{epsilon}
\epsilon &=& \dfrac{1}{2} \left(\dfrac{V_{,\phi}}{V}\right)^2,\\
\label{delta}
\delta &=& \dfrac{V_{\phi \phi}}{V}.
\end{eqnarray}

\bigskip
The number of e-foldings inside the slow-roll approximation is given by:

\begin{equation}
\label{N}
N \simeq \int^{\phi_\ini}_{\phi_\en}\dfrac{V}{V_{,\phi}} \diff\phi,
\end{equation}
where $\phi_\en$ is defined by $\epsilon\left(\phi_\en \right)$ when inflation ends.


\subsection{Solutions to the equations of motion}

\subsubsection{Numerical solution}
\begin{figure}[th!]
\includegraphics[scale=0.35]{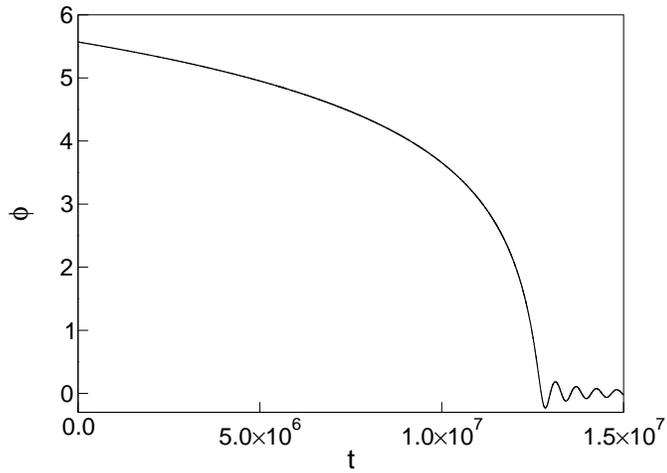}
\caption{Evolution of the scalar field $\phi$ for the Starobinsky inflationary model.}
\label{phi_ex}
\end{figure}

\begin{figure}[th!]
\vspace{1cm}
\includegraphics[scale=0.35]{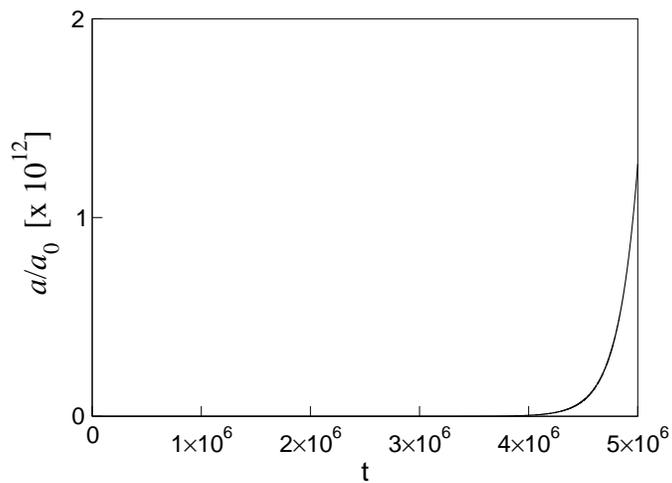}
\caption{Evolution of the scale factor $a$ for the Starobinsky inflationary model.}
\label{a_ex}
\end{figure}

The equations of motion \eqref{Friedmann} and \eqref{continuity} are solved numerically using the Software Mathematica\textsuperscript{\tiny\textregistered} version 12.0. Solving this system of coupled differential equations we obtain the behaviour of the scalar field and the scale factor with the cosmic time.
In Fig. \ref{phi_ex} we can observe the evolution of the scalar field $\phi$, when the scalar field starts to oscillate inflation ends,   this occurs at $t=1.27 \times 10^7$.
In Fig. \ref{a_ex} we can observe the evolution of the scale factor $a$ with time.

\subsubsection{Slow-roll approximation}

For the Starobinsky potential, the slow-roll equations. \eqref{Friedmannsr} and \eqref{continuitysr} becomes

\begin{eqnarray}
\label{Starobinsky_Friedmann_sr}
H&\simeq=& \dfrac{M^2}{\sqrt{3}}  \left(1-e^{-\sqrt{\sfrac{2}{3}}\,\phi}\right),\\
\label{Starobinsky_continuity_sr}
3 H \dot{\phi} &\simeq& -2 \sqrt{\dfrac{2}{3}} M^2 \left(1-e^{-\sqrt{\sfrac{2}{3}}\,\phi}\right).
\end{eqnarray}

\bigskip
From equations \eqref{Starobinsky_Friedmann_sr} and \eqref{Starobinsky_continuity_sr} we can obtain the expressions for the scale factor and the scalar field inside the slow-roll approximation

\begin{eqnarray}
\nonumber
a_\sr(t) &\simeq& \mathrm{Exp}\left[\dfrac{M^2t}{\sqrt{3}}-\dfrac{3}{4} \ln\left(e^{\sqrt{\sfrac{2}{3}} \, \phi_\ini}\right)\right.\\
&+&\left.\dfrac{3}{4} \ln\left(e^{\sqrt{\sfrac{2}{3}} \, \phi_\ini}-\dfrac{4 M^2 t}{3 \sqrt{3}}\right)\right],\\
\phi_\sr(t) &\simeq& \sqrt{\dfrac{3}{2}} \ln\left[\dfrac{1}{9}\left(e^{\sqrt{\sfrac{2}{3}} \, \phi_\ini}-4\sqrt{3} M^2 t \right)\right].
\end{eqnarray}

\begin{figure}[th!]
\includegraphics[scale=0.35]{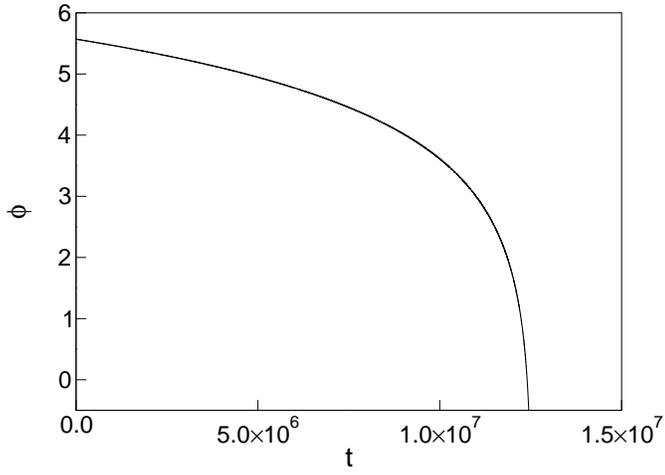}
\caption{Evolution of the scalar field $\phi$ for the Starobinsky inflationary model inside the slow-roll approximation.}
\label{phi_sr}
\end{figure}

\begin{figure}[th!]
\vspace{1cm}
\includegraphics[scale=0.35]{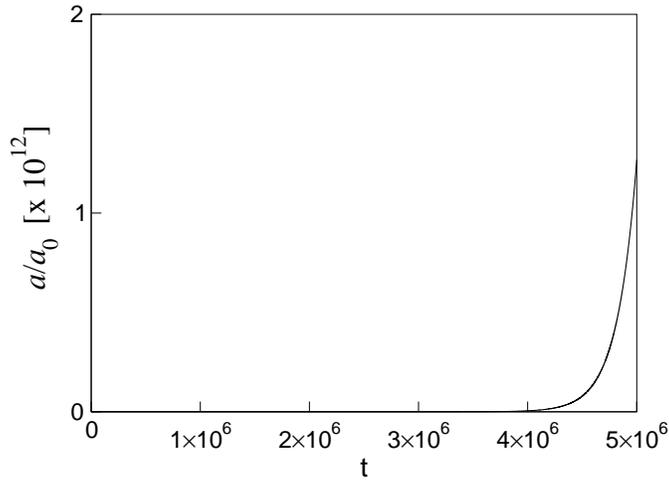}
\caption{Evolution of the scale factor $a$ for the Starobinsky inflationary model inside the slow-roll approximation.}
\label{a_sr}
\end{figure}

\bigskip
The behaviour of the scalar field $\phi$ and the scale factor $a$ inside the slow-roll approximation are shown in Fig. \ref{phi_sr} and \ref{a_sr}, respectively.

\bigskip
According to Eqs. \eqref{epsilon} and \eqref{delta} the slow-roll parameters for the Starobinsky inflationary model are given by \cite{adam:2019,samart:2019}:

\begin{eqnarray}
\label{epsilon_Starobinsky}
\epsilon &\simeq& \dfrac{4}{3 \left(e^{\sqrt{\sfrac{2}{3}} \, \phi} - 1 \right)^2},\\
\label{delta_Starobinsky}
\delta &\simeq& - \dfrac{4}{3} \dfrac{\left(e^{\sqrt{\sfrac{2}{3}} \, \phi} - 2 \right)^2} {\left(e^{\sqrt{\sfrac{2}{3}} \, \phi} - 1 \right)^2}.
\end{eqnarray}

\bigskip
At the end of inflation $\epsilon=1$, then from Eq. \eqref{epsilon_Starobinsky} we obtain that $\phi_{\en}=0.940$. The number of e-foldings are given by \cite{samart:2019}

\begin{equation}
\label{Nsr}
N_\sr=\dfrac{1}{4} \left[ 3 e^{\sqrt{\sfrac{2}{3}}\,\phi_\ini}-3 e^{\sqrt{\sfrac{2}{3}}\,\phi_\en} +\sqrt{6} \left(\phi_\en-\phi_\ini\right)\right].
\end{equation}

\bigskip
To solve the fine-tuning problems it is required  $N\geq 60$, from Eq. \eqref{Nsr} we obtain that $\phi_\ini \geq 5.57$.

\section{Equation of perturbations}

Since the scale factor $a$ and the field $\phi$ exhibit a simpler form in the physical time $t$ than in the conformal time $\eta$, we proceed to write the equations for the scalar  perturbations in the variable $t$. The relation between $t$ and $\eta$ is given via the equation $\D t= a\D \eta$. In this case, the equation for the perturbations can be written as

\begin{equation}
\label{dotu}
\ddot{u_k}+\frac{\dot{a}}{a}\dot{u_k}+\frac{1}{a^2}\left[k^2-\frac{\left(\dot{a}\dot{z_\sca}+a\ddot{z_\sca}\right)a}{z_\sca} \right]u_k=0,
\end{equation}
where $z_\sca=\sfrac{a \dot{\phi}}{H}$.

In order to apply the phase-integral approximation, we eliminate the terms $\dot{u}_k$ in Eq. \eqref{dotu}. We make the change of variables $u_k(t)=\frac{U_k(t)}{\sqrt{a}}$, obtaining that $U_{k}$ satisfy the differential equation:

\begin{equation}
\label{ddotUk}
\ddot{U}_k+R_\sca(k,t)U_k=0,
\end{equation}
with

\begin{equation}
\label{RS}
R_\sca(k,t)=\frac{1}{a^2}\left[k^2-\frac{\left(\dot{a}\dot{z_\sca}+a\ddot{z_\sca}\right)a}{z_\sca} \right]+\frac{1}{4a^2}\left(a^2-2a\ddot{a}\right),
\end{equation}

\medskip
\noindent
where $R_\sca(k,t)$ is  calculated numerically and $U(k)$ satisfies the asymptotic conditions

\begin{eqnarray}
\label{ceroUk}
U_k&\rightarrow&A_k \sqrt{a(t)} z_\sca(t),\quad  k\,t\rightarrow \infty,\\
\label{bordeUk}
U_k&\rightarrow&\sqrt{\frac{a(t)}{2k}}\exp{\left[-ik\eta(t)\right]}, \quad k\,t\rightarrow 0.
\end{eqnarray}

\begin{figure}[th!]
\begin{center}
\includegraphics[scale=0.35]{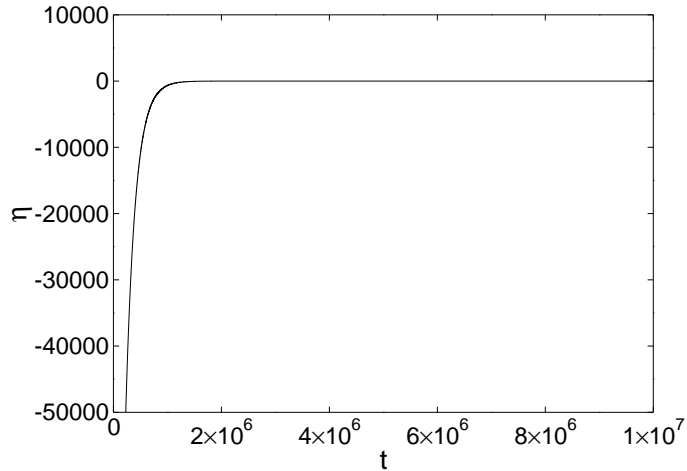}
\caption{\small{Behaviour of $\eta$ as a function of the physical time $t$ for the Starobinsky inflationary model.}}
\label{eta}
\end{center}
\end{figure}

In order to apply  the asymptotic condition \eqref{bordeUk}, we  use the relation between $\eta$ and $t$, which is given by:

\begin{equation}
\D\eta=\int_{t_\ini}^t \frac{\D t}{a(t)},
\end{equation}
where  $t_\ini= 1.27 \times 10^7$, so that $\eta$ is zero at the end of the inflationary epoch. The dependence  of $\eta$ on $t$ is shown in Fig. \ref{eta}. 

\medskip

From Fig. \ref{eta} we can observe that,

\begin{eqnarray}
\textnormal{when} \quad -k\,\eta\rightarrow 0          &\Rightarrow&  k\,t \rightarrow \infty,\\
\textnormal{when} \quad -k\,\eta\rightarrow \infty  &\Rightarrow& k\,t \rightarrow 0.
\end{eqnarray}
Eq. \eqref{dotu}, does not  possess an exact analytical solution. In order to solve the differential equation governing the scalar perturbations in the physical time $t$,  we  solve this equation numerically, then we use the slow-roll approximation, the uniform approximation,  the third-order phase integral approximation, and compare this results with the numerical calculation.

In order to solve the equation of perturbations with the semiclassical methods we have done a fit  to the numerical data for the scalar field and the scale factor. We have obtained the   following expressions

\begin{eqnarray}
\nonumber
a_\textnormal{fit}&=&2.71828^{(-3.41091 + 5.65 \times 10^{-6} t)} (94.4326 - 7.53333 \times 10^{-6} t)^{\sfrac{3}{4}},\\\\
\phi_\textnormal{fit}&=&1.23235 \ln\left(91.8173- 7.24641 \times 10^{-6}  t\right).
\end{eqnarray}

Once the mode equations for scalar  perturbations is
solved for different values of $k$, the power spectrum for scalar modes is given by the expression \cite{habib:2005b}

\begin{equation}
\label{PS}
P_S(k)= \lim_{-k\eta\rightarrow 0} \frac{k^3}{2 \pi^2}\left|\frac{u_k(\eta)}{z_{S}(\eta)} \right|^2.
\end{equation}

The power spectrum is usually fitted as a power-law $P_\sca (k) \propto k^{n_\sca -1}$, where $n_\sca$  corresponds to the spectral index.

\subsection{Solutions of the perturbation equation}
\subsubsection{Slow-roll approximation}

The scalar  power spectra in the slow-roll approximation to second-order is given by the expression \cite{stewart:2001}

\begin{eqnarray}
\label{PS_sr}
\nonumber
P_\sca^{\sr}(k)&\simeq&\left[1+(4b-2)\epsilon_1+2b\delta_1+\left(3b^2+2b-22\right.\right.\\
\nonumber
&+&\left.\frac{29\pi^2}{12}\right)\epsilon_1\delta_1+\left(3b^2-4+\frac{5\pi^2}{12}\right)\delta_1^2\\
&+&\left.\left(-b^2+\frac{\pi^2}{12}\right)\delta_2\right]\left.\left(\frac{H}{2\pi}\right)^2\left(\frac{H}{\dot{\phi}}\right)^2\right|_{k=aH} ,
\end{eqnarray}
where  $b$ is the Euler constant, $2-\ln2-b\simeq 0.7296$, $\ln2+b-1\simeq 0.2704$, and

\begin{eqnarray}
\epsilon_1&=&\frac{\dot{H}}{H^2},\\
\delta_1&=&\frac{1}{H\dot{\phi}}\frac{\D^2\phi}{\D t^2},\\
\delta_2&=&\frac{1}{H^2\dot{\phi}}\frac{\D^3\phi}{\D t^3}.
\end{eqnarray}

The spectral index in the slow-roll approximation is

\begin{eqnarray}
\label{nS_sr}
\nonumber
n_\sca^{\sr}(k)&\simeq&1-4\epsilon_1-2\delta_1+(8c-8)\epsilon_1^2+(10c-6)\epsilon_1\delta_1.\\
\end{eqnarray}

The expression  (\ref{PS_sr}) depends explicitly on time. In order to compute the scalar power spectrum we need to obtain the dependence  on the variable $k$. For a given value of $k$ ($0.0001\, \textnormal{Mpc}^{-1}  \leq k \leq 10\, \textnormal{Mpc}^{-1}  $) we obtain  $t_*$ from the relation $k=aH$. 

\subsubsection{Uniform approximation}

We want to obtain an approximate solution to the differential equation (\ref{ddotUk})  in the range where  $Q_\sca^2(k,t)$ have a simple root  at $t_\ret=\upsilon_\sca$, so that $Q_{\sca}^2(k,t)>0$ for  $0<t<t_\ret$ and $Q_{\sca}^2(k,t)<0$ for  $t>t_\ret$ as depicted in Fig. \ref{QSa}. Using the uniform approximation method \cite{berry:1972,habib:2002,rojas:2007b,rojas:2007c,rojas:2009}, we obtain that for  $0<t<t_\ret$ 

\begin{eqnarray}
\label{Ukzero}
\nonumber
U_k(k,t)&=&\left[\frac{\rho_\ele(k,t)}{Q_\sca^2(k,t)} \right]^{1/4} \left\{C_1
A_i[-\rho_\ele(k,t)]\right.\\
&+&\left. C_2 B_i[-\rho_\ele(k,t)] \right\},\\
\frac{2}{3}\left[\rho_\ele(k,t)\right]^{3/2}&=&\int_{t}^{t_\ret} \left[Q_{\sca}^2(k,t)\right]^{1/2}\D t,
\end{eqnarray}
\medskip
where  $C_1$ and  $C_2$ are two constants to be determined with the help of the boundary conditions (\ref{bordeUk}). 
For $t> t_\ret$

\begin{eqnarray}
\label{Ukinfinity}
\nonumber
U_k(k,t)&=&\left[\frac{-\rho_\ere(k,t)}{Q_\sca^2(k,t)} \right]^{1/4} \left\{C_1 A_i[\rho_\ere(k,t)]\right.\\
&+&\left. C_2 B_i[\rho_\ere(k,t)] \right\},\\
\frac{2}{3}\left[\rho_\ere(k,t)\right]^{3/2}&=&\int_{t_\ret}^{t} \left[-Q_{\sca}^2(k,t)\right]^{1/2}\D t,
\end{eqnarray}

For the computation of the power spectrum we need to take the limit $k\,t\rightarrow \infty$ of the solution  (\ref{Ukinfinity}). In this limit we have

\begin{eqnarray}
\label{limit_uk}
\nonumber
u_k^\ua(t)&\rightarrow&  \frac{C}{\sqrt{2\,a(t)}}\left[-Q_\sca^2(k,t)\right]^{-1/2}\\
\nonumber
&\times& \left\{ \frac{1}{2}\exp\left(-\int_{\upsilon_\sca}^{t}\left[-Q_\sca^2(k,t)\right]^{1/2} \D t\right)\right.\\
&+&\left.\im\,\exp\left(\int_{\upsilon_\sca}^{t}\left[-Q_\sca^2(k,t)\right]^{1/2} \D t\right)\right\},
\end{eqnarray}
where $C$ is a phase factor.  Using  the growing part  of the solutions   (\ref{limit_uk}),  one can compute the scalar  power spectrum using the uniform approximation method,

\begin{eqnarray}
P_\sca(k)&=&\lim_{-k t\rightarrow \infty} \frac{k^3}{2\pi^2} \left|\frac{u_k^\ua(t)}{z_\sca(t)}\right|^2.
\end{eqnarray} 


\subsubsection{Phase-integral approximation}
\begin{figure}[htbp]
\begin{center}
\subfigure[]{
\label{QSa}
\includegraphics[scale=0.3]{QS.eps}}
\subfigure[]{
\label{QSb}
\includegraphics[scale=0.3]{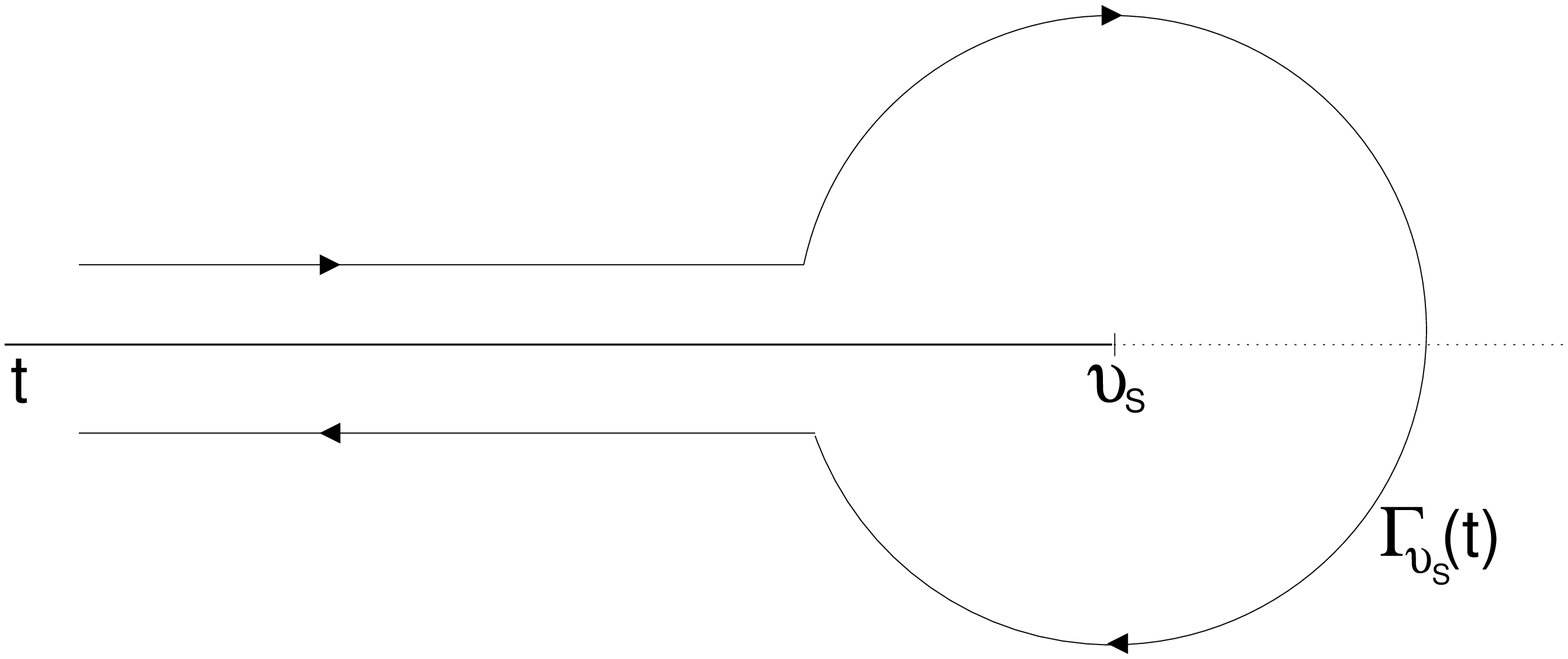}}
\subfigure[]{
\label{QSc}
\includegraphics[scale=0.3]{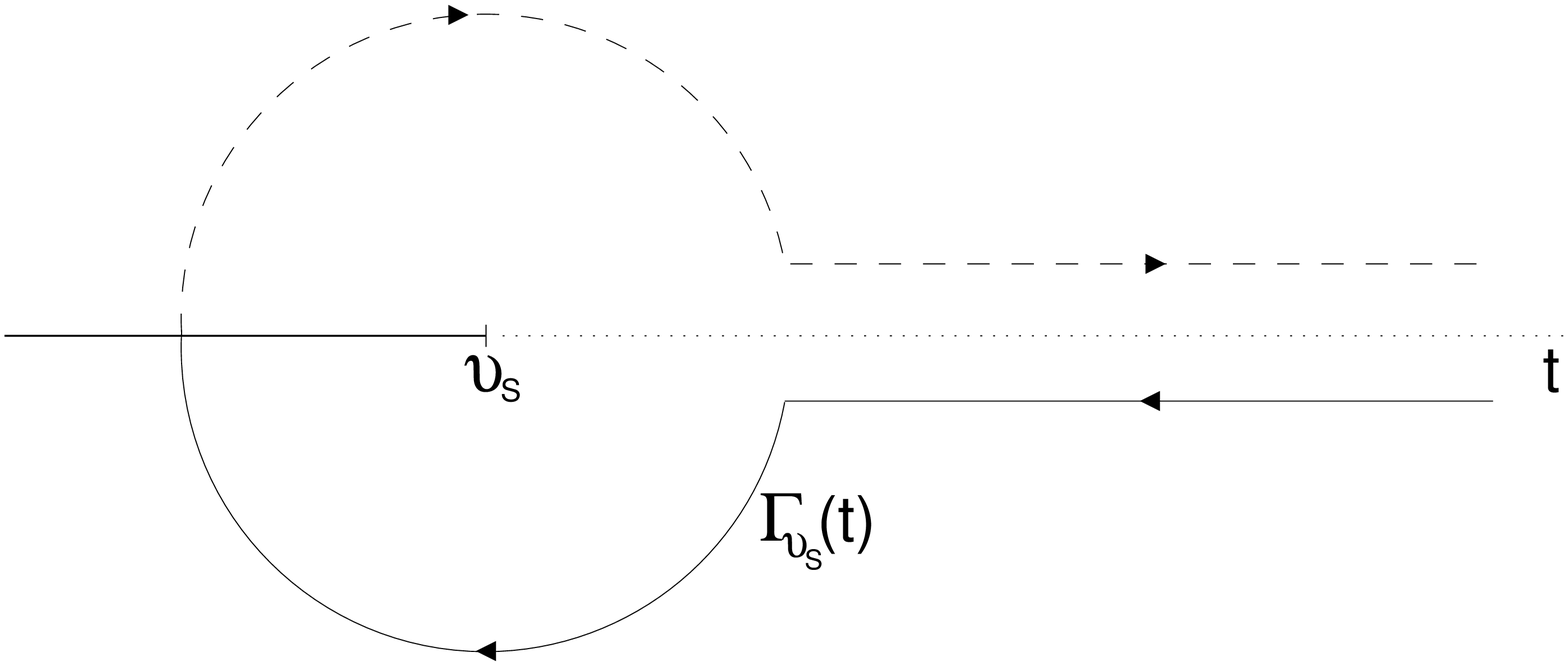}}
\caption{\small{(a) Behaviour of  $Q_\sca^2(k,t)$.
(b) Contour of integration $\Gamma_{\nu_\sca}(t)$ for $0<t<\nu_\sca$. 
(c) Contour of integration  $\Gamma_{\nu_\sca}(t)$ for $t>\nu_\sca$. 
The dashed lined indicates the part of the path on the second Riemann sheet.}}
	\end{center}
\end{figure}

In order to solve Eq. (\ref{ddotUk})  with the help of the phase-integral approximation \cite{froman:1996}, we choose the following base functions $Q_\sca$ for the scalar  perturbations

\begin{eqnarray}
\label{Q}
Q_\sca^2(k,t)&=&R_\sca(k,t),
\end{eqnarray}
where  $R_\sca(k,t)$ is given by  Eq. (\ref{RS}).  Using this selection, the phase-integral approximation is  valid as  $k t\rightarrow \infty$, limit where we should impose the condition (\ref{ceroUk}), where the validity condition   $\mu \ll 1$ holds.  The bases functions  $Q_\sca(k,t)$  possess turning points, for each mode $k$ this turning points represent the horizon. There are two ranges  where to define the solution. To the left of the turning point  $0<t<t_\ret$ we have the classically permitted region  $Q_{\sca}^2(k,t)>0$ and to the right of the turning point $t>t_\ret$ corresponding to the classically forbidden region $Q_{\sca}^2(k,t)<0$, such as it is shown in Figs \ref{QSa}.

The mode  $k$ equations for the scalar  perturbations (\ref{ddotUk})  in the phase-integral approximation has two solutions: 

 For $0<t <t_\ret$

\begin{eqnarray}
\label{ukleft}
\nonumber
u^\phai_k(t)&=& \frac{c_1}{\sqrt{a(t)}}\left|q_\sca^{-1/2}(k,t)\right| \cos{\left[\left|\omega_\sca(k,t)\right|-\frac{\pi}{4}\right]} \\
&+& \frac{c_2}{\sqrt{a(t)}}\left|q_\sca^{-1/2}(k,t)\right| \cos{\left[\left|\omega_\sca(k,t)\right|+\frac{\pi}{4}\right]}.
\end{eqnarray}

For $t>t_\ret$

\begin{eqnarray}
\label{ukright}
\nonumber
u^\phai_k(t)&=&\frac{c_1}{2\sqrt{a(t)}}\left|q_\sca^{-1/2}(k,t)\right|\exp\left[-\left|\omega_\sca(k,t)\right|\right]\\
&+& \frac{c_2}{\sqrt{a(t)}} \left|q_\sca^{-1/2}(k,t)\right| \exp\left[\left|\omega_\sca(k,t)\right|\right].
\end{eqnarray}

Using the phase-integral approximation up to third order ($2N+1=3\rightarrow N=1$), we have that $q_{\sca}(k,t)$  can be expanded in the form

\begin{eqnarray}
q_\sca(k,t)&=&\sum_{n=0}^1 Y_{2n_\sca}(k,t) Q_\sca(k,t)\\
\label{q1}
&=&\left[Y_{0_\sca}(k,t)+Y_{2_\sca}(k,t)\right] Q_\sca(k,t).
\end{eqnarray}

In order to compute $q_\sca(k,t)$, we  compute $Y_{2_\sca}(k,t)$, and the required functions  $\varepsilon_{0_\sca}(k,t)$.  The expression  (\ref{q1}) gives  a third-order approximation for $q_\sca(k,t)$.  In order to compute $\omega_\sca(k,t)$  we make a contour integration following the path indicated in Fig. \ref{QSb}-(c).

\begin{eqnarray}
\omega_\sca(k,t)&=&\omega_{0_\sca}(k,t)+ \omega_{2_\sca}(k,t),\\
&=&\int_{\upsilon_\sca}^{t}Q_\sca(k,t)\D t+\frac{1}{2}\int_{\Gamma_{\upsilon_\sca}}Y_{2_\sca}(k,t)Q_\sca(k,t)\D t,\\
&=&\int_{\upsilon_\sca}^{t}Q_\sca(k,t)\D t+\frac{1}{2}\int_{\Gamma_{\upsilon_\sca}}f_{2_\sca}(k,t)\D t,
\end{eqnarray}
\medskip
where

\begin{eqnarray}
f_{2n_\sca}(k,t)&=&Y_{2_\sca}(k,t)Q_\sca(k,t).
\end{eqnarray}
The functions $f_{2_\sca}(k,t)$  have the following functional dependence:

\begin{eqnarray}
\label{A}
f_{2_\sca}(k,t)&=&A(k,t)(t-\upsilon_\sca)^{-5/2},
\end{eqnarray}
where $A(k,t)$ is regular at $\upsilon_\sca$ . With the help of  the function  (\ref{A}) we compute the integrals for $\omega_{2n}$  up to $N=1$ using the contour indicated in  Fig. \ref{QSb}-(c). The expressions for  $\omega_{2n}$ permit one to obtain the third-order phase integral approximation of the solution to the equation for scalar perturbations (\ref{ddotUk}).  The constants $c_1$ and $c_2$ are obtained using the limit  $k\,t\rightarrow 0$ of the solutions on the left side of the turning point (\ref{ukleft}), and  are given by the expressions

\begin{eqnarray}
c_1&=&-\im\,c_2,\\
c_2&=&\frac{\e^{-\im\frac{\pi}{4}}}{\sqrt{2}}\e^{-\im\left[k\,\eta(0)+\left|\omega_{0_\sca}(k,0)\right|\right]}.
\end{eqnarray}

\medskip
In order to compute the scalar power spectrum, we need to calculate the limit as  $k\,t\rightarrow \infty$ of the growing part of the solutions on the right side of the turning point  given by   Eq. (\ref{ukright}) for scalar perturbations.

\begin{eqnarray}
P_\sca(k)&=&\lim_{-k t\rightarrow \infty} \frac{k^3}{2\pi^2} \left|\frac{u_k^\phai(t)}{z_\sca(t)}\right|^2.
\end{eqnarray}

\section{Results}

\begin{figure}[th!]
\includegraphics[scale=0.35]{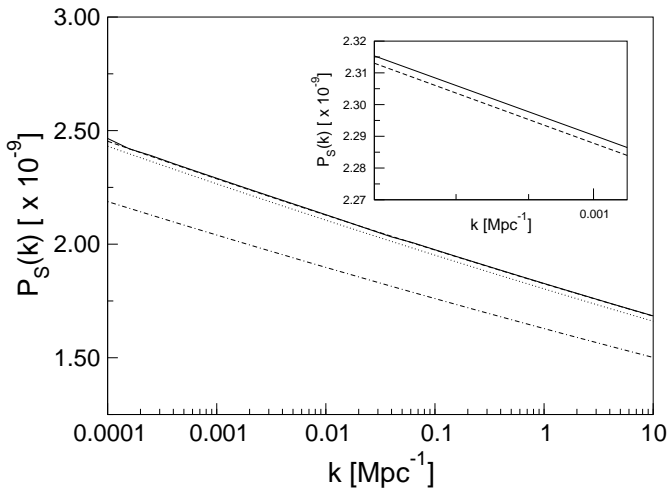}
\caption{$P_\sca(k)$  for the Starobinsky inflationary model.  Solid line: numerical result; dashed line: third-order phase-integral approximation; dot-dashed line: uniform approximation, dotted line: second-order slow-roll.  The inset is an enlargement of the figure.}
\label{PS_sca}
\end{figure}

We have calculated the scalar power spectrum for the Starobinsky inflationary model, with the second-order slow-roll approximation, the uniform-approximation and, the phase-integral method up to third-order of approximation. Figure \ref{PS_sca} shows $P_\sca (k)$ using each method of approximation. We can observe that the semiclassical methods work very well and they are of easiest implementation that the numerical method.

\begin{figure}[th!]
\includegraphics[scale=0.35]{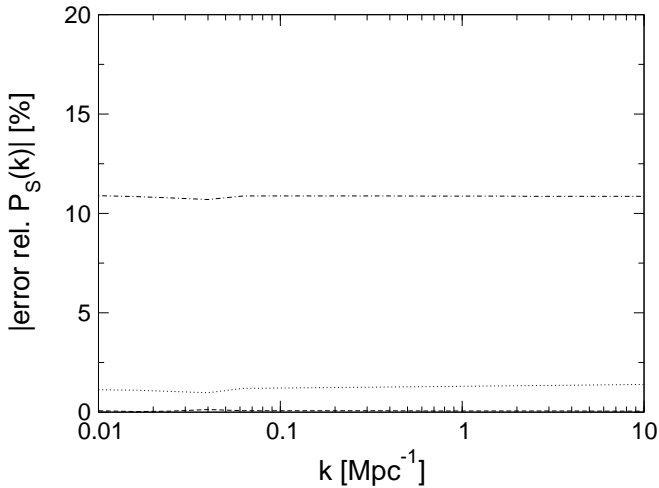}
\caption{Relative error with respect to the numerical result of $P_\sca(k)$   for the Starobinsky inflationary model.  Dashed line: third-order phase-integral approximation; dot-dashed line: uniform approximation, dotted line: second-order slow-roll approximation.}
\label{error_PS}
\end{figure}

Fig. \ref{error_PS} shows the relative error of each  approximation method with respect to the numerical result. The uniform approximation deviates from the numerical result in $11\%$, the second-order slow roll approximation deviates in $1.39\%$, whereas that the phase-integral approximation up to third-order reduces to $0.28\%$.

Doing a fitting of the power spectrum to a power-law form $P_\sca(k) \propto k^{n_\sca-1}$ 
the value of the spectral index for each approximation method is given in table \ref{nS}. 
The value of the spectral index according to the Planck result is $n_\sca=0.960 \pm 0.007$ \cite{kehagias:2014}, then our results are according to the observational data.

\begin{table}
\begin{tabular}{ccc}
\toprule
Method                           & $n_\sca$  & rel. err (\%)\\ 
\midrule
Numerical                      & $\;\;0.967351$     & $\;\;$              \\
Second-order slow roll               & $\;\;0.967103$       &  $\;\;0.025637$   \\
Uniform approximation & $\;\;0.967505$       & $\;\; 0.015920$ \\
Phase integral method   & $\;\;0.967486$    & $\;\;0.013956$   \\
\bottomrule
\end{tabular}
\caption{Value of  $n_\sca$ obtained  with different approximation methods for the Starobinsky inflationary model.}
\label{nS}
\end{table}

\section{Conclusions}

As in our previous work \cite{rojas:2007b,rojas:2007c,rojas:2009,rojas:2012} we have showed that the semiclassical methods are useful tools to calculate  the scalar power spectrum and the spectral index for different inflationary models. We have obtained for the Starobinsky inflationary model a relative error of the order of $11\%$ with the uniform approximation method, a relative error of order $1.39\%$ for the second-order slow-roll approximation, and a relative error of order $0.28\%$ with the phase integral method up to third-order in approximation. Also the spectral index is according to the results of Planck $2018$ \cite{akrami:2018}.


\section{Acknolegment}

The authors want to express their gratitude to Professor Alexei Starobinsky for useful discussions.


\bibliographystyle{unsrt}
\bibliography{starobinsky}

\begin{thebibliography}{10}

\bibitem{guth:1981}
A.~H. Guth.
\newblock {Inflationary universe: A possible solution to the horizon and
  flatness problems}.
\newblock {\em Phys. Rev. D}, 23:347, 1981.

\bibitem{guth:1982}
A.~H. Guth and So-Young Pi.
\newblock {Fluctuations in the new inflationary universe}.
\newblock {\em Phys. Rev. Lett.}, 49:1110, 1983.

\bibitem{liddle:2000}
A.~R. Liddle and D.~H. Lyth.
\newblock {\em {Cosmological inflation and large-scale structure}}.
\newblock Cambridge University Press, 2000.

\bibitem{martin:2014}
C.~Ringeval J.~Martin and V.~Vennin.
\newblock {Encyclopaedia Inflationaris}.
\newblock {\em Phys. Dark Univ.}, 5-6:75--235, 2014.

\bibitem{akrami:2018}
Y.~Akrami \textit{et al.}
\newblock {Planck 2018 results. X. Constraints on inflation}.
\newblock {\em arXiv:1807.06211}, 2018.

\bibitem{barrow:1988a}
{John D. Barrow}.
\newblock {The Premature Recollapse Problem in Closed Inflationary Universes}.
\newblock {\em Nucl. Phys. B}, 296:697, 1988.

\bibitem{barrow:1988b}
{John D. Barrow and S. Cotsakis}.
\newblock {Inflation and the Conformal Structure of Higher Order Gravity
  Theories}.
\newblock {\em Phys. Lett. B}, 214:515, 1988.

\bibitem{starobinsky:1980}
A.~A. Starobinsky.
\newblock {A new type of isotropic cosmological models without singularity}.
\newblock {\em Phys. Lett. B}, 91:99, 1980.

\bibitem{linde:2014}
A.~Linde.
\newblock {Inflationary Cosmology after Planck 2013}.
\newblock {\em arXiv:1402.0526}, 2014.

\bibitem{diValentino:2017}
E.~Di Valentino and L.~Mersini-Houghton.
\newblock {Testing predictions of the quantum landscape multiverse 1: the
  Starobinsky inflationary potential}.
\newblock {\em JCAP}, 2, 2017.

\bibitem{paliathanasis:2017}
A.~Paliathanasis.
\newblock {Analytic solution of the Starobinsky model for inflation}.
\newblock {\em Eur. Phys. J C}, 77:438, 2017.

\bibitem{adam:2019}
C.~Adam and D.~Varela.
\newblock {The superpotential method in cosmological inflation}.
\newblock {\em arXiv:1901}, 2019.

\bibitem{granada:2019}
L.~N. Granada and D.~F. Jimenez.
\newblock {Slow-roll inflation with exponential potential in scalar-tensor
  models}.
\newblock {\em Eur. Phys. J. C}, 79:772, 2019.

\bibitem{samart:2019}
D.~Samart and P.~Channuie.
\newblock {Unification of inflation and dark matter in the Higgs-Starobinsky
  model}.
\newblock {\em Eur. Phys. J. C}, 79:347, 2019.

\bibitem{chowdhury:2019}
C.~Ringeval D.~Chowdhury, J.~Martin and V.~Vennin.
\newblock {Inflation after Planck: Judgment Day}.
\newblock {\em arXiv:1902.03951}, 2019.

\bibitem{habib:2002}
S.~Habib, A.~Heinen, K.~Heitmann, G.~Jungman, and C.~Molina-Par\'is.
\newblock {The Inflationary Perturbation Spectrum}.
\newblock {\em Phys. Rev. Lett.}, 89:281301, 2002.

\bibitem{casadio:2005}
{R. Casadio, F. Finelli, M. Luzzi, and G. Venturi}.
\newblock {Improved WKB analysis of cosmological perturbations}.
\newblock {\em Phys. Rev. D}, 71:043517, 2005.

\bibitem{casadio:2006}
{R. Casadio, F. Finelli, A. Kamenshchik, M. Luzzi, and G. Venturi}.
\newblock {The method of comparison equations for cosmological perturbations}.
\newblock {\em JCAP}, 04:011, 2006.

\bibitem{rojas:2007b}
Clara Rojas and V\'ictor~M. Villalba.
\newblock {Computation of inflationary cosmological perturbations in the
  power-law inflatioary model using the phase-integral method}.
\newblock {\em Phys. Rev. D}, 75:063518, 2007.

\bibitem{rojas:2007c}
V\'ictor~M. Villalba and Clara Rojas.
\newblock {Applications of the phase integral method ins ome inflationary
  scenarios}.
\newblock {\em J. Phys. Conf. Ser.}, 66:012034, 2007.

\bibitem{rojas:2009}
Clara Rojas and V\'ictor~M. Villalba.
\newblock {Computation of inflationary cosmological perturbations in chaotic
  inflationary scenarios using the phase-integral method}.
\newblock {\em Phys. Rev. D}, 79:103502, 2009.

\bibitem{rojas:2012}
Clara Rojas and V\'ictor~M. Villalba.
\newblock {Computation of the power spectrum in chaotic
  $\frac{1}{4}\lambda\phi^4$ inflation}.
\newblock {\em JCAP}, 003:1, 2012.

\bibitem{martin:2019}
J.~Martin.
\newblock {Cosmic Inflation: Trick or Treat?}
\newblock {\em arXiv:1902.05286}, 2019.

\bibitem{mishra:2018}
V.~Sahni S.~S.~Mishra and A.~V. Toporensky.
\newblock {Initial conditions for inflation in an FRW universe}.
\newblock {\em Phys. Rev. D}, 98:083538, 2018.

\bibitem{habib:2005b}
S.~Habib, A.~Heinen, K.~Heitmann, and G.~Jungman.
\newblock {Inflationary Perturbations and Precision Cosmology}.
\newblock {\em Phys. Rev. D}, 71:043518, 2005.

\bibitem{stewart:2001}
E.~D Stewart and J.~Gong.
\newblock {The density perturbation power spectrum to second-order corrections
  in the slow-roll expansion}.
\newblock {\em Phys. Lett. B}, 510:1, 2001.

\bibitem{berry:1972}
M.~Berry and K.~E. MounT.
\newblock {Semiclassical Approximations in Wave Mechanics}.
\newblock {\em Rep. Prog. Phys.}, 35:315, 1972.

\bibitem{froman:1996}
N.~Fr\"oman and P.~O. F\"oman.
\newblock {\em {Phase-Integral Method. Allowing Nearlying Transition Point}},
  volume~40.
\newblock Springer Tracts in Natural Philosophy, 1996.

\bibitem{kehagias:2014}
A.~M.~Dizgah A.~Kehagias and A.~Riotto.
\newblock {Remarks on the Starobinsky model of inflation and its descendants}.
\newblock {\em Phys. Rev. D}, 89:043527, 2014.

\end{thebibliography}
\end{document}